\documentclass[twocolumn,showpacs,preprintnumbers,amsmath,amssymb,floatfix]{revtex4}

\usepackage{verbatim}
\usepackage{graphicx}
\usepackage{dcolumn}
\usepackage{bm}
\usepackage{color}
\usepackage{url}
\usepackage{amsmath}
\usepackage{color}

\graphicspath{{./figures/}}
\newcommand{\be}{\begin{equation}}
\newcommand{\ee}{\end{equation}}
\newcommand{\ba}{\begin{eqnarray}}
\newcommand{\ea}{\end{eqnarray}}
\newcommand{\nn}{\nonumber}

\newcommand{\hyp}{\mathcal{H}}
\def\ltsima{$\; \buildrel < \over \sim \;$}
\def\simlt{\lower.5ex\hbox{\ltsima}}
\def\gtsima{$\; \buildrel > \over \sim \;$}
\def\simgt{\lower.5ex\hbox{\gtsima}}
\DeclareMathOperator{\nonGR}{non-GR}

\begin{document}

\title[blah]{Testing the no-hair theorem with black hole ringdowns using TIGER}

\author{J.~Meidam$^{1}$, M.~Agathos$^{1}$, C.~Van Den Broeck$^{1}$, J.~Veitch$^{1,2}$, and B.S.~Sathyaprakash$^{3}$}
\affiliation{$^{1}$Nikhef -- National Institute for Subatomic Physics, Science Park 105, 1098 XG Amsterdam, The Netherlands\\
$^2$School of Physics and Astronomy, University of Birmingham, Edgbaston, Birmingham B15 2TT, United Kingdom\\
$^3$School of Physics and Astronomy, Cardiff University, 5, The Parade, Cardiff, CF24 3AA, United Kingdom}
\date{\today}

\begin{abstract}
The Einstein Telescope (ET), a proposed third-generation gravitational wave observatory, would enable tests of the no-hair theorem by looking at the characteristic frequencies and damping times of black hole ringdown signals. In previous work it was shown that with a single $500 - 1000\,M_\odot$ black hole at distance $\lesssim 6$ Gpc (or redshift $z \lesssim 1$), deviations of a few percent in the  frequencies and damping times of dominant and sub-dominant modes would be within the range of detectability. Given that such sources may be relatively rare, it is of interest to see how well the no-hair theorem can be tested with events at much larger distances and with smaller signal-to-noise ratios, thus accessing a far bigger volume of space and a larger number of sources. We employ a model selection scheme called TIGER (Test Infrastructure for GEneral Relativity), which was originally developed to test general relativity with weak binary coalescence signals that will be seen in second-generation detectors such as Advanced LIGO and Advanced Virgo. TIGER is well-suited for the regime of low signal-to-noise ratio, and information from a population of sources can be combined so as to arrive at a stronger test. By performing a range of simulations using the expected noise power spectral density of Einstein Telescope, we show that with TIGER, similar deviations from the no-hair theorem as considered in previous work will be detectable with great confidence using $\mathcal{O}(10)$ sources distributed uniformly in co-moving volume out to 50 Gpc ($z \lesssim 5$).
\end{abstract}


\maketitle

\section{Introduction}\label{sec:intro}

The no-hair theorem states that a black hole that has settled down to its final stationary vacuum state is determined only by its mass, spin and electric charge \cite{Israel1967,Israel1968,Hawking1971,Hawking1972,Carter1970}. Astrophysical black holes are thought to be electrically neutral so that only mass and spin need to be considered, leading to the Kerr geometry. When a black hole is formed as a result of the inspiral and merger of two other compact objects, it will undergo `ringdown' as it evolves towards its quiescent state. This process can be modeled by considering linear perturbations of the Kerr metric, or quasi-normal modes, which are characterized by frequencies $\omega_{lm}$ and damping times $\tau_{lm}$ \cite{Ruffini1971,Leaver1985,Kokkotas1999,Berti2009}. Since the underlying Kerr spacetime is only characterized by its mass $M$ and spin $J$, these frequencies and times are constrained by linearized general relativity to only depend on these quantities through specific functional relationships, so that observational tests of these dependences would constitute a test of the no-hair theorem, and hence of general relativity (GR);\footnote{For the purposes of this paper we will treat the no-hair theorem as if it were a prediction of GR, so that a violation of the theorem implies a GR violation. However, it should be noted that existing proofs require strong assumptions, such as analyticity of spacetime; see \cite{Chrusciel1994} for an overview. For this reason, the no-hair theorem is often referred to as the no-hair conjecture.} this was first hinted at by Detweiler \cite{Detweiler1980}, made concrete by Dreyer \emph{et al.}~\cite{Dreyer2004}, and further explored  in \cite{Berti2006,Berti2007,Kamaretsos2012a}.

Recently Gossan, Veitch, and Sathyaprakash \cite{Gossan2012} investigated the possibility of performing this kind of test using Einstein Telescope (ET), a proposed third-generation ground-based gravitational wave detector \cite{ET}, as well as with the space-based \emph{e}LISA \cite{eLISA}. These authors evaluated two methods for checking the dependences $\omega_{lm} = \omega_{lm}^{\rm GR}(M,J)$ and $\tau_{lm} = \tau_{lm}^{\rm GR}(M,J)$ predicted by GR: Bayesian parameter estimation and model selection. Specifically, one can write  possible deviations from these dependences as
\ba
\omega_{lm} &=& \omega_{lm}^{\rm GR}(M,J)\,(1 + \delta\hat{\omega}_{lm}), \label{omega}\\
\tau_{lm} &=& \tau_{lm}^{\rm GR}(M,J)\,(1 + \delta\hat{\tau}_{lm}), \label{tau}
\ea
and then (a) calculate how well the dimensionless quantities $\delta\hat{\omega}_{lm}$, $\delta\hat{\tau}_{lm}$ can be measured, or (b) compare the evidences for two models: one where the $\delta\hat{\omega}_{lm}$, $\delta\hat{\tau}_{lm}$ are free parameters, and another in which they are all identically zero, corresponding to the GR prediction. In practice, the authors of \cite{Gossan2012} restricted their attention to the set
\be
\{\delta\hat{\omega}_{22}, \delta\hat{\omega}_{33}, \delta\hat{\tau}_{22}\}. \label{domega22domega33dtau22}
\ee
It was found that for black holes with masses in the range $500 - 1000\,M_\odot$ at a distance of 6 Gpc,  ET would allow for measurements of $\delta\hat{\omega}_{22}$, $\delta\hat{\omega}_{33}$, and $\delta\hat{\tau}_{22}$ with accuracies of a few percent for the first two parameters, and about 10\% for the third. (For comparison, boson stars in the same mass range would cause $\delta\hat{\omega}_{22}$ and $\delta\hat{\tau}_{22}$ to be of order 1 \cite{Yoshida1994}.) With model selection and assuming a $500\,M_\odot$ black hole, a deviation of a few percent in $\delta\hat{\omega}_{22}$ could be discriminated from GR with $\ln B^{\rm dev}_{\rm GR} > 10$, were $B^{\rm dev}_{\rm GR}$ is the \emph{Bayes factor}, or ratio of evidences, for the model that deviates from GR (with the variables in Eq.~(\ref{domega22domega33dtau22}) as extra free parameters) versus the GR model.

How frequently might one test GR in this way? Coalescence rates of intermediate-mass binary black holes which would give rise to ringdowns with masses in the above range are highly uncertain \cite{Miller2004,Fregeau2006,Amaro-Seoane2006}; ET may see between a few and a few thousands per year \cite{Amaro-Seoane2010,Gair2011}. Gossan \emph{et al.}~considered single, relatively loud sources, but one will also want to combine information from multiple, possibly weak signals out to large distances so as to maximally exploit the available set of detections. Since deviations from the no-hair theorem may be such that $\delta\hat{\omega}_{22}$, $\delta\hat{\omega}_{33}$, and/or $\delta\hat{\tau}_{22}$ take on different non-zero values for different sources, when doing parameter estimation it will not be possible to combine posterior probability densities from multiple events unless one already assumes GR to be correct. On the other hand, although Bayesian model selection does lend itself quite easily to the utilization of all available detections, if one lets  $\{\delta\hat{\omega}_{22}, \delta\hat{\omega}_{33}, \delta\hat{\tau}_{22}\}$ (and possibly more of the $\delta\hat{\omega}_{lm}$, $\delta\hat{\tau}_{lm}$) vary all at the same time, one may be penalized if the corresponding model is insufficiently parsimonious, \emph{i.e.}~if the correct model involves a smaller number of additional parameters.

In \cite{Li2012a,Li2012b,VanDenBroeck2013,Agathos2013,Agathos2014}, a more general algorithm for testing GR was developed, called TIGER (Test Infrastructure for GEneral Relativity). Take a gravitational waveform model as predicted by GR, and introduce deformations parameterized by dimensionless quantities $\delta\xi_i$, $i = 1, 2, \ldots, N_T$ such that all of the $\delta\xi_i$ being zero corresponds to GR being correct. One can then ask the question: ``Do one or more of the $\delta\xi_i$ differ from zero?" Let us denote the corresponding hypothesis by $\hyp_{\rm modGR}$, and the GR hypothesis by $\hyp_{\rm GR}$. Now, there is no waveform model that corresponds to $\hyp_{\rm modGR}$. However, as shown in \cite{Li2012a}, one can define logically disjoint `sub-hypotheses' $H_{i_1 i_2 \ldots i_k}$, in each of which a \emph{fixed} set of parameters $\{\delta\xi_{i_1}, \delta\xi_{i_2}, \ldots, \delta\xi_{i_k} \}$ are non-zero while $\delta\xi_j = 0$ for $j \notin \{i_1,  i_2, \ldots, i_k\}$. There are $2^{N_T} - 1$ such sub-hypotheses, corresponding to the non-empty sub-sets of the full set $\{\delta\xi_1, \delta\xi_2, \ldots, \delta\xi_{N_T}\}$. The $H_{i_1 i_2 \ldots i_k}$ \emph{do} have waveform models associated with them that can be compared with the data, and $\hyp_{\rm modGR}$ can be expressed as the logical union of all the sub-hypotheses:
\be
\hyp_{\rm modGR} = \bigvee_{i_1 < i_2 < \ldots < i_k;\, k \leq N_T} H_{i_1 i_2 \ldots i_k}.
\label{union}
\ee
Given a catalog of detections $d_1, d_2, \ldots, d_\mathcal{N}$ and whatever background information $I$ one may possess, one can then compute the \emph{odds ratio} for $\hyp_{\rm modGR}$ against $\hyp_{\rm GR}$:
\ba
\mathcal{O}^{\rm modGR}_{\rm GR} &\equiv& \frac{P(\hyp_{\rm modGR}|d, I)}{P(\hyp_{\rm GR}|d, I)} \nn\\
&=& \frac{\alpha}{2^{N_T} - 1} \sum_{i_1 < i_2 < \ldots < i_k; k \leq N_T} \prod_{A=1}^\mathcal{N} {}^{(A)}B^{i_1 i_2 \ldots i_k}_{\rm GR}. \nn\\
\label{odds}
\ea
Here $\alpha$ is an unimportant scaling factor which below will be set to unity, and the Bayes factors ${}^{(A)}B^{i_1 i_2 \ldots i_k}_{\rm GR}$ for a detection $d_A$ are given by
\be
{}^{(A)}B^{i_1 i_2 \ldots i_k}_{\rm GR} \equiv \frac{P(d_A|H_{i_1 i_2 \ldots i_k}, I)}{P(d_A|\hyp_{\rm GR}, I)},
\ee
with $P(d_A|H_{i_1 i_2 \ldots i_k}, I)$ and $P(d_A|\hyp_{\rm GR}, I)$ the \emph{evidences} for $H_{i_1 i_2 \ldots i_k}$ and $\hyp_{\rm GR}$, respectively. For basic assumptions and detailed derivations we refer to \cite{Li2012a,Li2012b,VanDenBroeck2013}.

The TIGER formalism has been evaluated extensively in the context of binary neutron star inspirals that will be observed by second-generation detectors such as Advanced LIGO \cite{aLIGO}, Advanced Virgo \cite{AdV}, GEO-HF \cite{GEO}, KAGRA \cite{KAGRA}, and LIGO-India \cite{IndIGO}. In \cite{Li2012a,Li2012b} it was shown that, thanks to the introduction of the $H_{i_1 i_2 \ldots i_k}$, the method avoids potential problems due to insufficient parsimony, is well-suited to dealing with weak signals, and enables the discovery of a wide range of deviations from GR, including ones that are well outside the particular parameterized waveform family used; moreover, information from multiple sources can trivially be combined.

However, TIGER is not tied to any particular gravitational waveform model and can be applied to testing the no-hair theorem with ringdown signals. Consider the $N_T = 3$ testing parameters of \cite{Gossan2012},
\be
\delta\xi_1 = \delta\hat{\omega}_{22}, \,\,\,\,\,\delta\xi_2 = \delta\hat{\omega}_{33}, \,\,\,\,\,\delta\xi_3 = \delta\hat{\tau}_{22}.
\label{domegadtau}
\ee
$\hyp_{\rm modGR}$, the hypothesis that one or more of the $\delta\xi_i$ deviate from their GR value, is then the logical union of $2^3-1 = 7$ sub-hypotheses $H_1$, $H_2$, $H_3$, $H_{12}$, $H_{13}$, $H_{23}$, and $H_{123}$. Here $H_1$ is the hypothesis that $\delta\xi_1 \neq 0$ while $\delta\xi_2 = \delta\xi_3 = 0$, $H_{13}$ the hypothesis that both $\delta\xi_1 \neq 0$ and $\delta\xi_3 \neq 0$ but $\delta\xi_2 = 0$, and similarly for the other sub-hypotheses. In the above language, the model selection set-up of Gossan \emph{et al.}~\cite{Gossan2012} only involved calculating, for a single source, the Bayes factor
\be
B^{123}_{\rm GR} = \frac{P(d|H_{123}, I)}{P(d|\hyp_{\rm GR}, I)}.
\ee
It would be of great interest to see how our ability to discern violations of the no-hair theorem with ringdown signals would improve if  the full formalism of TIGER were brought to bear. This will be the main topic of the present paper.

When evaluating the odds ratio $\mathcal{O}^{\rm modGR}_{\rm GR}$ of Eq.~(\ref{odds}) using one or more detected signals, we may find that there is no reason to believe that GR is incorrect. However, in that case it will still be of interest to \emph{measure} $\delta\hat{\omega}_{22}$, $\delta\hat{\omega}_{33}$, and $\delta\hat{\tau}_{22}$ for each source and combine the resulting posterior density distributions so as to arrive at a joint result for the entire catalog of detections. This we will also do, and as we shall see, potentially tight constraints can be set on these parameters.

This paper is structured as follows. In Sec.~\ref{sec:SignalAndAnalysis} we explain our assumptions regarding Einstein Telescope as well as our waveform models for signals and templates, and the set-up of the simulations. In Sec.~\ref{sec:TestNoHair} we evaluate TIGER's ability to  perform tests of the no-hair theorem. The possibility of precision measurements of the free parameters in case we have no reason to doubt GR is discussed in Sec.~\ref{sec:PDFs}. Sec.~\ref{sec:Conclusions} provides a summary and conclusions.

Throughout this paper we will use units such that $G = c = 1$.

\section{Detectors, waveform models, and set-up of the simulations}
\label{sec:SignalAndAnalysis}

\subsection{Einstein Telescope}

In 2011, a conceptual design study for Einstein Telescope was concluded \cite{ET,Punturo2010,Sathyaprakash2011,Sathyaprakash2012}. ET is envisaged to consist of three underground detectors arranged in a unilateral triangle with 10 km sides. Each detector is composed of two interferometers: a cryogenic one with improved sensitivity at low frequencies (up to $\sim 40$ Hz) due to the suppression of thermal noise, and a non-cryogenic interferometer which is more sensitive at high frequencies (up to several kHz) due to higher laser power, which reduces quantum noise.

\begin{figure}[!htp]
\includegraphics[width=\columnwidth]{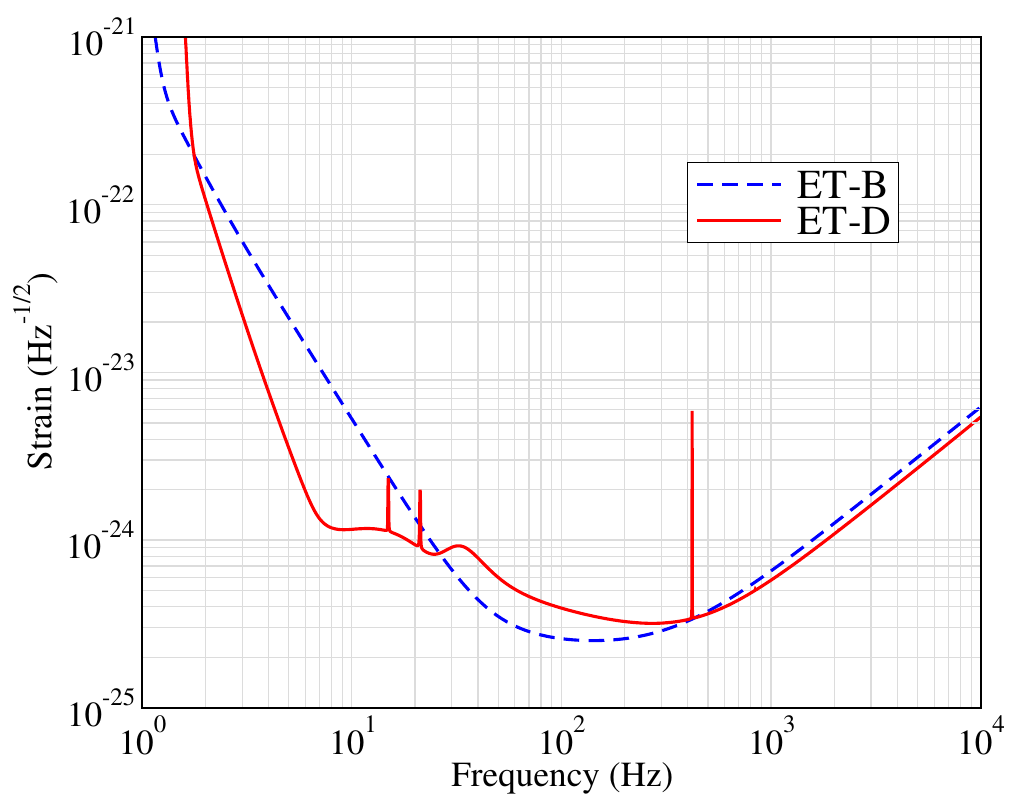}
\caption{The strain sensitivity of Einstein Telescope as envisaged in \cite{ET}, labeled ET-D, compared with the older ET-B curve.}
\label{fig:etd}
\end{figure}

The combined strain sensitivity as a function of frequency for each detector is the one labeled `ET-D' in Fig.~\ref{fig:etd} \cite{Hild2010}. In the same plot we show the older `ET-B' curve used by Gossan \emph{et al.}~\cite{Gossan2012}, where only a single interferometer was assumed for each of the detectors. Currently ET-D corresponds to the most detailed assessment of the possible noise budget of Einstein Telescope. We will take the lower cut-off frequency to be 10 Hz, and the dominant mode frequencies considered in this paper will roughly lie between 15 and  100 Hz, a range in which, for the most part, ET-D is less sensitive than ET-B, by up to a factor of 2. We note that in reality it may be possible to achieve a lower frequency cut-off of only a few Hz, and between there and $~25$ Hz, ET-D is actually more sensitive than ET-B, which would lead to comparably better visibility of higher-mass sources; hence our assumptions are conservative. For each of the three detectors, stretches of simulated stationary, Gaussian noise were produced with ET-D as  underlying power spectral density. Simulated signals were added coherently to each of the three data streams, taking into account the different detector responses \cite{Regimbau2012}.

\subsection{Waveform models}

The  ringdown signal is given by a superposition of quasi-normal modes characterized by triples of integers $(l, m, n)$, where $l \geq 2$ and $m = -l, -l + 1, \ldots, l - 1, l$; $n \geq 0$ is an overtone index \cite{Teukolsky1972,Teukolsky1973,Kokkotas1999,Berti2009}. Here we will only consider the modes with $n = 0$, as overtones with $n > 0$ are not significantly excited and have much shorter damping times \cite{Berti2006}. The `plus' and `cross' polarizations read
\ba
h_+(t) &=& \frac{M}{D_{\rm L}} \sum_{l,m>0} A_{l|m|}e^{-t/\tau_{lm}} Y_+^{lm}(\iota) \cos(\omega_{lm}t - m \phi), \nn\\
h_{\times}(t) &=& -\frac{M}{D_{\rm L}} \sum_{l,m>0} A_{l|m|}e^{-t/\tau_{lm}} Y_{\times}^{lm}(\iota) \sin(\omega_{lm}t - m \phi), \nn\\ \label{hpluscross}
\ea
where the $Y_+^{lm}(\iota)$, $Y_\times^{lm}(\iota)$ can be written in terms of spin-weighted spherical harmonics of weight $-2$:
\ba
Y_+^{lm}(\iota)       &\equiv& {}_{-2}Y^{lm}(\iota,0) + (-1)^l {}_{-2}Y^{l,-m}(\iota,0),\label{Yplm} \nn\\
Y_{\times}^{lm}(\iota)&\equiv& {}_{-2}Y^{lm}(\iota,0) - (-1)^l {}_{-2}Y^{l,-m}(\iota,0).\label{Yclm}
\ea
In the above, $M$ is the observed mass of the black hole, which is related to the intrinsic mass by $M = (1 + z)\,M_{\rm intr}$, with $z$ the redshift; $D_{\rm L}$ is the luminosity distance to the source; $\iota$ is the angle between the black hole's spin and the line of sight; and $\phi$ is the azimuth angle of the black hole with respect to the observer. 
%Note that in principle there will be additional phase offsets $\phi_{lm}$ in (\ref{hpluscross}), which we expect to be identical to each other for non-spinning progenitor binaries; although we will also consider spinning progenitors, in the absence of predictions for their dependence on progenitor parameters we set them to zero, as was also done in \cite{Gossan2012,Kamaretsos2012c}. 
Note that in principle there will be additional phase offsets $\phi_{lm}$ in (\ref{hpluscross}); since analytic fits for their dependence on progenitor parameters are not yet available, we set them to zero, as was also done in \cite{Gossan2012,Kamaretsos2012c}. 
$\omega_{lm}(M,j)$ and $\tau_{lm}(M,j)$ are the characteristic frequencies and damping times of the modes, respectively, as functions of the mass and of the dimensionless spin $j = J/M^2$.

As in \cite{Gossan2012}, we only consider the modes $(l,m) = (2,2)$, $(2,1)$, $(3,3)$, $(4,4)$, which are among the most dominant ones. Analytic expressions for the mode amplitudes $A_{l|m|}$ are not available, but there exist accurate fits to numerical simulations. The authors of \cite{Gossan2012} took the \emph{progenitor} black holes to be non-spinning, in which case one can use the approximate expressions for the $A_{l|m|}$ in terms of the symmetric mass ratio $\nu = m_1 m_2/(m_1 + m_2)^2$ (with $m_1$, $m_2$ the progenitor component masses)  from Kamaretsos \emph{et al.}~\cite{Kamaretsos2012a}.\footnote{More recent fits for the $A_{l|m|}$ in the case of non-spinning progenitors can be found in \cite{London2014}.} Here we will relax this assumption and include the effect of non-zero progenitor spins in the waveforms, using more recent results. For spinning progenitors, Kamaretsos, Hannam, and Sathyaprakash \cite{Kamaretsos2012b} found that mainly $A_{21}$ is strongly affected, and a good fit for all the relevant amplitudes is given by
\ba
A_{22}(\nu) &=& 0.864 \nu,\label{A22}\\
A_{21}(\nu) &=& 0.43\,\left[ \sqrt{1-4\nu} - \chi_{\rm eff} \right] A_{22}(\nu),\label{A21}\\
A_{33}(\nu) &=& 0.44(1-4\nu)^{0.45} A_{22}(\nu).\label{A33}\\
A_{44}(\nu) &=& \left[ 5.4 (\nu - 0.22)^2 + 0.04 \right] A_{22}(\nu),\label{A44}
\ea
where
\be
\chi_{\rm eff} = \frac{1}{2}\left( \sqrt{1-4\nu}\,\chi_1 + \chi_- \right), \label{chiEff}
\ee
with
\be
\chi_- = \frac{m_1\chi_1 - m_2\chi_2}{M_{\rm in}}. \label{chiMinus}
\ee
Here $(m_1, m_2)$ and $(\chi_1, \chi_2)$ are, respectively, the progenitor component masses and dimensionless spin magnitudes, and $M_{\rm in}$ is the initial total mass of the system, which to reasonable approximation we can take to be equal to the mass of the final black hole.

For the frequencies $\omega_{lm}$ and damping times $\tau_{lm}$ there also exist good fits, which can be expressed through the \emph{quality factors} $Q_{lm} = \omega_{lm} \tau_{lm}/2$:
\ba
M\omega &=& f_1 + f_2 (1 - j)^{f_3},\\
Q &=& q_1 + q_2 (1 - j)^{q_3},
\ea
where for the values of the coefficients $f_1$, $f_2$, $f_3$, $q_1$, $q_2$, $q_3$ we refer to \cite{Berti2006}. Finally, there exists a simple fit for the spin $j$ of the final black hole in terms of the component masses $(m_1, m_2)$ and spins $(\vec{\chi}_1, \vec{\chi}_2)$ \cite{Rezzolla2008,Barausse2009}, for which we refer to \cite{Barausse2009}.

For the simulated sigals, or \emph{injections}, we choose progenitor spins $\vec{\chi}_1$, $\vec{\chi}_2$ from a distribution with isotropic directions, and a Gaussian distribution for the magnitudes centered on 0.7, with standard deviation 0.2 and hard cut-offs at 0.5 and 0.99 \cite{Berti2008}; note that the value of 0.7 roughly corresponds to what one gets from the coalescence of non-spinning, equal mass binary black holes. The mass $M$ is drawn from a uniform distribution between 500 and 1000 $M_\odot$, and the mass ratio $q = m_1/m_2$ from a uniform distribution between 0.3 and 1. Amplitudes are computed as in Eqs.~(\ref{A22})-(\ref{chiMinus}), where we take $\chi_{1,2} = |\vec{\chi}_{1,2}|$, and the final spin $j$ is calculated from component masses and spins using the formula of Barausse and Rezzolla \cite{Barausse2009}. With these choices for masses and spins, the characteristic frequency $f_{22} = \omega_{22}/(2\pi)$ of the dominant ringdown mode ranges from about 15 to 100 Hz, while the inspiral signal, which ends roughly at $f_{\rm LSO}  = (6^{3/2}\pi M)^{-1}$, stays below the lower cut-off frequency of 10 Hz and hence is never in the sensitive frequency band.  Redshifts are taken to be between 1.5 and 5, and sources are placed uniformly in co-moving volume assuming a $\Lambda$CDM cosmology with $(\Omega_{\rm M}, \Omega_\Lambda, h_0) = (0.27, 0.73, 0.70)$, so that luminosity distances approximately range from 10 to 50 Gpc. Since part of the exercise is to stress-test the TIGER framework, we only analyze sources with signal-to-noise ratio (SNR) $< 30$, corresponding to a minimum angle-averaged distance of 14.97 Gpc ($z = 1.90$). Sky positions $(\theta, \varphi)$ and orientations $(\iota, \psi)$ are drawn from uniform distributions on the sphere. To gauge our sensitivity to deviations in $\omega_{22}(M, j)$, $\omega_{33}(M, j)$, and $\tau_{22}(M, j)$, we introduce constant relative shifts $\delta\hat{\omega}_{22}$, $\delta\hat{\omega}_{33}$, and $\delta\hat{\tau}_{22}$ as explained in the introduction.

For the \emph{templates}, we only take $\chi_{\rm eff}$ and $j$ to be the spin-related free parameters, as the progenitor component spins $\vec{\chi}_1$ and $\vec{\chi}_2$ will not be separately measurable from a ringdown signal alone.\footnote{The progenitor spins may become  measurable if more modes are included than the ones considered here.} The free parameters for the waveform model corresponding to the GR hypothesis $\hyp_{\rm GR}$ are then
\be
\vec{\theta}_{\rm GR} = \left\{ M, \nu, j, \chi_{\rm eff}, D_{\rm L}, \theta, \varphi, \psi, \iota, \phi, t_0 \right\},
\ee
where $t_0$ is the time of arrival of the signal at the detector. The prior on $M$ is chosen to be uniform between 300 and 1200 $M_\odot$, and the one for the symmetric mass ratio $\nu$ is flat between 0.01 and 0.25; in terms of the mass ratio $q = m_1/m_2$ this range corresponds to $0.01 \lesssim q \leq 1$.  The prior on $j$ is uniform between 0.01 and 0.99, and the one on $\chi_{\rm eff}$ is uniform between $-1$ and 1. Sky positions and orientations are taken to be uniform on the sphere, and the prior on distance is uniform in co-moving volume between 8 and 60 Gpc. $t_0$ is taken to be in a window of width 100 ms.

\subsection{TIGER for ringdown}

To apply TIGER in the context of ringdown, we introduce the same parameterized deformations of the waveform as in \cite{Gossan2012}, namely the ones of Eqs.~(\ref{omega})-(\ref{domega22domega33dtau22}). The parameter spaces corresponding to the various sub-hypotheses $H_{i_1 i_2 \ldots i_k}$ of $\hyp_{\rm modGR}$ are given by
\ba
H_1 &\longleftrightarrow& \{\vec{\theta}_{\rm GR}, \delta\hat{\omega}_{22}\}, \nn\\
H_2 &\longleftrightarrow&  \{\vec{\theta}_{\rm GR}, \delta\hat{\omega}_{33}\}, \nn\\
H_3 &\longleftrightarrow&  \{\vec{\theta}_{\rm GR}, \delta\hat{\tau}_{22}\}, \nn\\
H_{12} &\longleftrightarrow& \{\vec{\theta}_{\rm GR}, \delta\hat{\omega}_{22},\delta\hat{\omega}_{33}\}, \nn\\
H_{13} &\longleftrightarrow& \{\vec{\theta}_{\rm GR}, \delta\hat{\omega}_{22}, \delta\hat{\tau}_{22}\},  \nn\\
H_{23} &\longleftrightarrow& \{\vec{\theta}_{\rm GR}, \delta\hat{\omega}_{33}, \delta\hat{\tau}_{22}\}, \nn\\
H_{123} &\longleftrightarrow&  \{\vec{\theta}_{\rm GR}, \delta\hat{\omega}_{22}, \delta\hat{\omega}_{33}, \delta\hat{\tau}_{22}\}. \nn\\
\ea
Given a detection $d_A$, the corresponding Bayes factors ${}^{(A)}B^1_{\rm GR}$, ${}^{(A)}B^2_{\rm GR}$, ${}^{(A)}B^3_{\rm GR}$, ${}^{(A)}B^{12}_{\rm GR}$, ${}^{(A)}B^{13}_{\rm GR}$, ${}^{(A)}B^{23}_{\rm GR}$, and ${}^{(A)}B^{123}_{\rm GR}$ are calculated using
\be
{}^{(A)}B^{i_1 i_2 \ldots i_k}_{\rm GR} = \frac{{}^{(A)}B^{i_1 i_2 \ldots i_k}_{\rm noise}}{{}^{(A)}B^{\rm GR}_{\rm noise}},
\label{Bfactors}
\ee
where ${}^{(A)}B^{i_1 i_2 \ldots i_k}_{\rm noise}$, ${}^{(A)}B^{\rm GR}_{\rm noise}$ are, respectively, the Bayes factors for $H_{i_1 i_2 \ldots i_k}$ and $\hyp_{\rm GR}$ against the hypothesis that the data contain only noise. The latter are computed using an appropriate adaptation of the nested sampling algorithm as implemented by Veitch and Vecchio \cite{Veitch2008a,Veitch2008b,Veitch2010}.

For completeness, we give the expression for the odds ratio $\mathcal{O}^{\rm modGR}_{\rm GR}$ in the present context; given a catalog of $\mathcal{N}$ ringdown signals, it reads
\ba
&& \mathcal{O}^{\rm modGR}_{\rm GR} \nn\\
&& = \frac{1}{7} \left[ \prod_{A=1}^\mathcal{N}{}^{(A)}B^1_{\rm GR} +  \prod_{A=1}^\mathcal{N}{}^{(A)}B^2_{\rm GR} + \prod_{A=1}^\mathcal{N}{}^{(A)}B^3_{\rm GR} \right.  \nn\\
&& \left. \,\,\,\,\,\,\,\,\,\,\,\,\,\,\,\,\, + \prod_{A=1}^\mathcal{N}{}^{(A)}B^{12}_{\rm GR} + \prod_{A=1}^\mathcal{N}{}^{(A)}B^{13}_{\rm GR} + \prod_{A=1}^\mathcal{N}{}^{(A)}B^{23}_{\rm GR} \right. \nn\\
&& \left. \,\,\,\,\,\,\,\,\,\,\,\,\,\,\,\,\, + \prod_{A=1}^\mathcal{N}{}^{(A)}B^{123}_{\rm GR} \right].
\ea

In practice it is often convenient to work with the \emph{logarithm} of the odds ratio, $\ln\mathcal{O}^{\rm modGR}_{\rm GR}$. If GR is correct, then naively one would expect $\mathcal{O}^{\rm modGR}_{\rm GR} < 1$, or $\ln\mathcal{O}^{\rm modGR}_{\rm GR} < 0$. However, features in the noise can have a detrimental effect on the measurement of the log odds ratio, and in practice one can obtain slightly positive values of $\ln\mathcal{O}^{\rm modGR}_{\rm GR}$ even if no deviation from GR is present. For this reason one usually constructs a \emph{background distribution} $P(\ln\mathcal{O}|\hyp_{\rm GR}, \kappa_{\rm GR}, I)$ \cite{Li2012a,Li2012b,VanDenBroeck2013,Agathos2013,Agathos2014}. Here $\kappa_{\rm GR}$ denotes a large number of (catalogs of) injections with GR waveforms, for each of which one computes $\ln\mathcal{O}^{\rm modGR}_{\rm GR}$, whose normalized distribution constitutes $P(\ln\mathcal{O}|\hyp_{\rm GR}, \kappa_{\rm GR}, I)$. Given a maximum false alarm probability $\beta$ that one is willing to tolerate, one can use this background to set a threshold $\ln\mathcal{O}_\beta$ for the \emph{measured} log odds ratio to overcome; this threshold is defined such that $\beta$ is the fraction of the background distribution that is above $\ln\mathcal{O}_\beta$:
\be
\beta = \int_{\ln\mathcal{O}_\beta}^\infty P(\ln\mathcal{O}|\hyp_{\rm GR}, \kappa_{\rm GR}, I)\,d\ln\mathcal{O}.
\ee
In reality there will only be a single value for the measured log odds ratio, computed from the signals one actually detects. However, if one wants to assess how likely it is that a particular type of deviation from GR, denoted by $\hyp_{\nonGR}$, will lead to a log odds ratio above threshold, then one can construct a \emph{foreground distribution} $P(\ln\mathcal{O}|\hyp_{\nonGR}, \kappa_{\nonGR}, I)$, where this time $\kappa_{\nonGR}$ is a set of injections whose waveforms are in accordance with the given GR violation. One can then define the \emph{efficiency} $\zeta$ as the fraction of the foreground that is above threshold:
\be
\zeta = \int_{\ln\mathcal{O}_\beta}^\infty P(\ln\mathcal{O}|\hyp_{\nonGR}, \kappa_{\nonGR}, I)\,d\ln\mathcal{O}.
\ee
This can be viewed as the probability that the particular kind of deviation from GR considered will be discovered with a false alarm probability of at most $\beta$.

In what follows, we will consider both the case where only a single ringdown detection is ever made by ET, so that $\mathcal{N} = 1$, and the case where multiple detections are made. As mentioned before, the event rate for ringdowns with mass in the range $500- 1000\,M_\odot$ is highly uncertain, but a few tens of detections out to tens of Gpc is consistent with expectations in the literature \cite{Miller2004,Fregeau2006,Amaro-Seoane2006,Amaro-Seoane2010,Gair2011}. Below we will show results where injections are randomly combined into catalogs of $\mathcal{O}(10)$ sources each.

To evaluate TIGER's ability to find deviations from the no-hair theorem, we will mostly study its behavior in the following cases:
\begin{enumerate}
\item There is a 10\% deviation in the dominant-mode frequency $\omega_{22}$, but other mode frequencies as well as the damping times are unaffected; \emph{i.e.}, the injections have $(\delta\hat{\omega}_{22}, \delta\hat{\omega}_{33}, \delta\hat{\tau}_{22}) = (0.1, 0, 0)$.
\item There is a 10\% deviation in the $(3,3)$ mode frequency $\omega_{33}$, but no deviation in other frequencies or in the damping times; \emph{i.e.}, $(\delta\hat{\omega}_{22}, \delta\hat{\omega}_{33}, \delta\hat{\tau}_{22}) = (0, 0.1, 0)$.
\item There is a 10\% deviation in the dominant-mode damping time $\tau_{22}$, but no deviation in other damping times or in the frequencies: $(\delta\hat{\omega}_{22}, \delta\hat{\omega}_{33}, \delta\hat{\tau}_{22}) = (0, 0, 0.1)$.
\item There is a 25\% deviation in $\tau_{22}$, but no deviation in other damping times or in the frequencies: $(\delta\hat{\omega}_{22}, \delta\hat{\omega}_{33}, \delta\hat{\tau}_{22}) = (0, 0, 0.25)$.
\end{enumerate}
Note that in the notation introduced above, this means that, in turn, we take $H_1$, $H_2$, and $H_3$ to be the correct hypotheses; the resulting distributions of log odds ratio for single sources as well as catalogs of sources will be our foreground distributions. We also consider the case where the no-hair theorem holds, \emph{i.e.}~$\hyp_{\rm GR}$ is the correct hypothesis and the injections have $\delta\hat{\omega}_{22} = \delta\hat{\omega}_{33} = \delta\hat{\tau}_{22} = 0$. The log odds ratio distributions resulting from the latter, again for single sources and catalogs of sources, will be our backgrounds.

\section{Testing the no-hair theorem with TIGER}
\label{sec:TestNoHair}

\begin{figure*}[!htp]
\centering
  \begin{tabular}{ccc}%{@{}ccc@{}}
    \includegraphics[width=.33\textwidth]{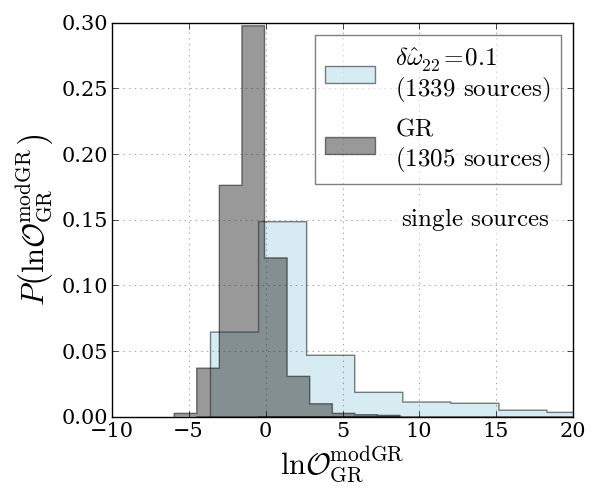} &
    \includegraphics[width=.33\textwidth]{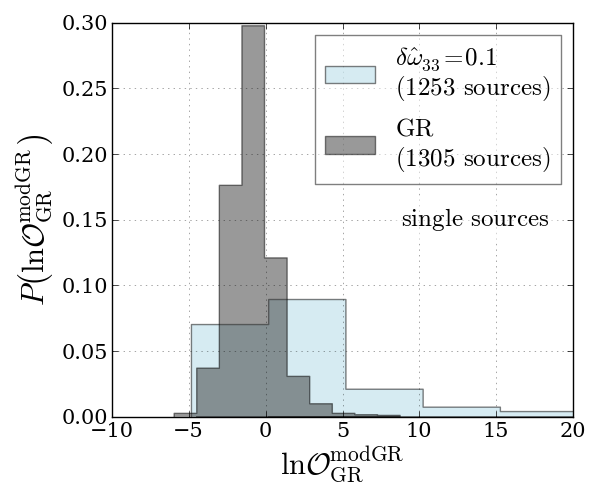} &
    \includegraphics[width=.33\textwidth]{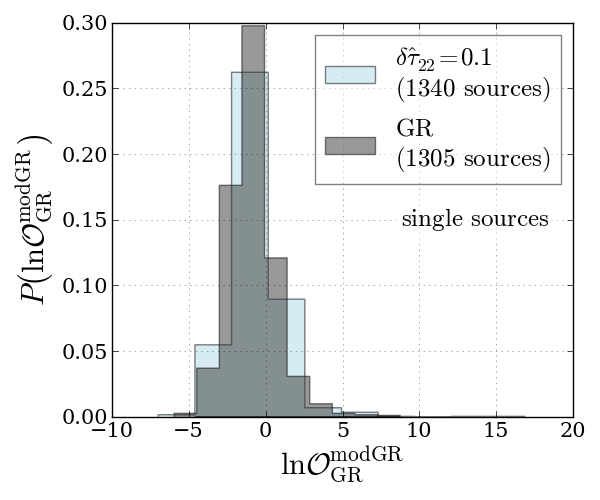}
  \end{tabular}
  \caption{Single-source GR background distributions (dark grey) and foreground distributions (light gray), for a 10\% deviation in $\omega_{22}$ (left), a 10\% deviation in $\omega_{33}$ (middle), and a 10\% deviation in $\tau_{22}$ (right). In all three cases there is significant overlap between background and foreground; for a maximum tolerable false alarm probability of $\beta = 0.05$, the efficiencies are, respectively, 47\%, 46\%, and 5\%.}
\label{fig:singlesources}
\end{figure*}

Let us first focus on background and foreground distributions for the case where only a single ringdown detection is ever made by ET, \emph{i.e.}~$\mathcal{N} = 1$. Results for $\sim 1300$ sources are shown in Fig.~\ref{fig:singlesources}. We see that in all of the cases, there is significant overlap between background and foreground, so that with a single source one has no guarantee that violations of the no-hair theorem at these levels will be picked up. With a maximum tolerable false alarm probability of $\beta = 0.05$, the efficiency $\zeta$ for a 10\% shift in $\omega_{22}$ is 0.47, and for a 10\% shift in $\omega_{33}$ it is 0.46. Note how the efficiencies for deviations in $\omega_{22}$ and $\omega_{33}$ are comparable; with our choice for the injected range of mass ratios ($0.3 < q < 1$, or $0.18 \lesssim \nu < 0.25$) there will be sources with $A_{33} > A_{22}$ as well as sources with $A_{33} < A_{22}$; see Fig.~1 of \cite{Gossan2012}. For a 10\% shift in $\tau_{22}$ we find $\zeta = 0.05$. Thus, even with only a single detection, one will have a reasonable chance of finding a GR violation of the given size in $\omega_{22}$ and $\omega_{33}$; however, the same shift in $\tau_{22}$ will be essentially unobservable.

\begin{figure*}[!htp]
\centering
  \begin{tabular}{ccc}%{@{}ccc@{}}
    \includegraphics[width=.33\textwidth]{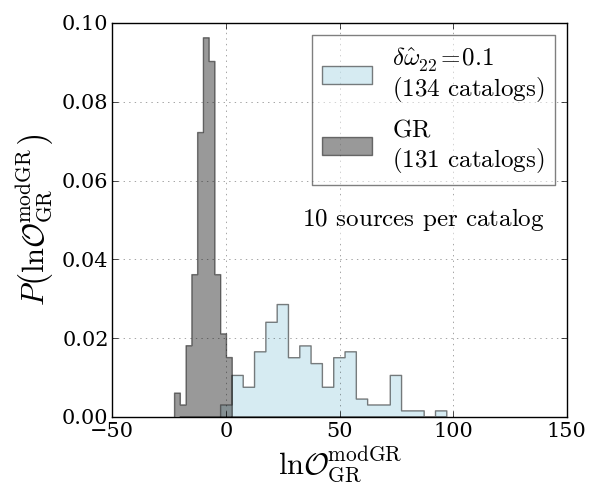} &
    \includegraphics[width=.33\textwidth]{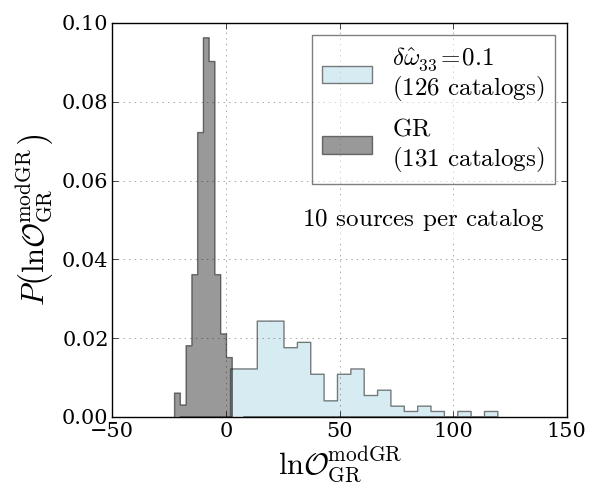} &
    \includegraphics[width=.33\textwidth]{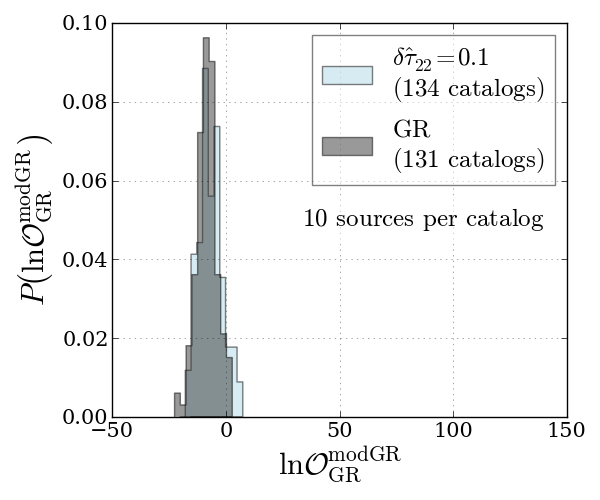}
  \end{tabular}
  \caption{GR background distributions (dark gray) and foreground distributions (light gray), for a 10\% deviation in $\omega_{22}$ (left), a 10\% deviation in $\omega_{33}$ (middle), and a 10\% deviation in $\tau_{22}$ (right). This time we considered \emph{catalogs} of 10 sources each. Again with $\beta = 0.05$, this time efficiencies of 98\% are attained for the two mode frequencies. On the other hand, the deviation in $\tau_{22}$ remains hard to detect, with an efficiency of only 14\%.}
\label{fig:catalogs}
\end{figure*}

At least for anomalies in $\omega_{22}$ and $\omega_{33}$, the situation changes dramatically if information from multiple detections can be combined. This is shown in Fig.~\ref{fig:catalogs}, for catalogs of 10 sources each. For the same maximum false alarm probability and the given shifts in $\omega_{22}$, $\omega_{33}$, and $\tau_{22}$, the efficiencies become, respectively, 0.98, 0.98, and 0.13. Thus, there is a very significant improvement in the first two cases, but the shift in $\tau_{22}$ remains hard to observe.

\begin{figure*}[!htp]
\centering
  \begin{tabular}{ccc}%{@{}ccc@{}}
    \includegraphics[width=.33\textwidth]{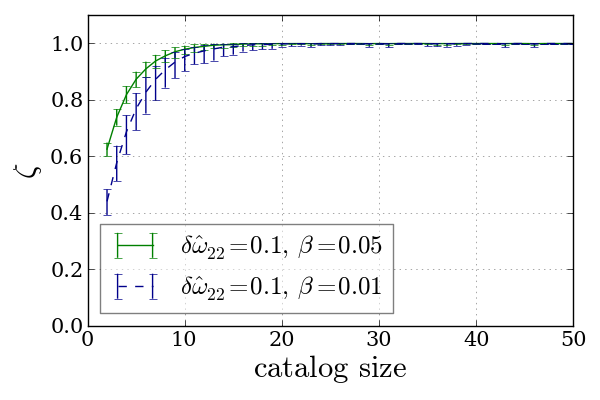} &
    \includegraphics[width=.33\textwidth]{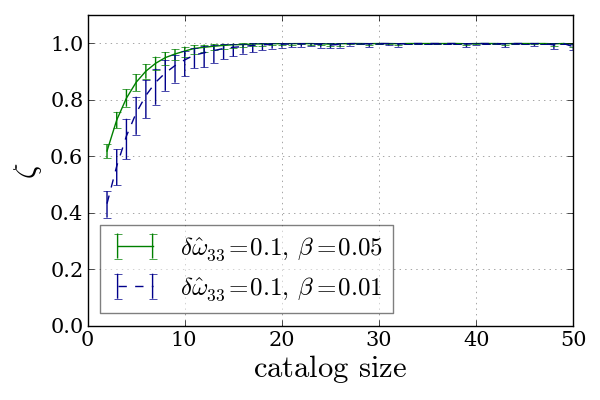} &
    \includegraphics[width=.33\textwidth]{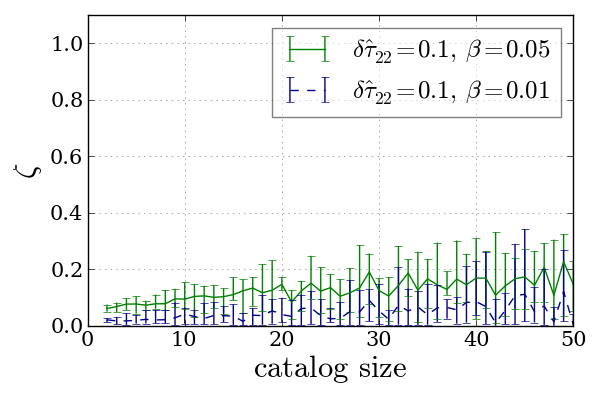}
  \end{tabular}
  \caption{Growth of the efficiency $\zeta$ with the number of sources per catalog, for a 10\% deviation in $\omega_{22}$ (left), a 10\% deviation in $\omega_{33}$ (middle), and a 10\% deviation in $\tau_{22}$ (right), for maximum tolerable false alarm probabilities $\beta = 0.05$ and $\beta = 0.01$, respectively. In order to understand uncertainties in $\zeta$ due to having a finite number of catalogs, the available simulated sources were randomly combined into catalogs to obtain 1000 different realizations. Shown are the median efficiencies (solid and dashed lines) and 95\% confidence intervals.}
\label{fig:efficiencies}
\end{figure*}

It is also of interest to see how the efficiencies grow with an increasing number of sources per catalog. This is shown in Fig.~\ref{fig:efficiencies}, for two choices of maximum tolerable false alarm probability: $\beta = 0.05$ and $\beta = 0.01$. Due to the finite number of catalogs considered, inevitably the numbers we quote for efficiencies are not exact; in the plot we show medians and 95\% confidence intervals obtained for $\zeta$ when randomly combining the available simulated sources into catalogs of a given size in 1000 different ways. For the cases $\delta\hat{\omega}_{22} = 0.1$ and $\delta\hat{\omega}_{33} = 0.1$, we see that for either value of $\beta$, the efficiency reaches essentially 100\% for $\sim 20$ sources per catalog. However, for a GR violation with $\delta\hat{\tau}_{22} = 0.1$ and as many as 50 sources per catalog, even with $\beta = 0.05$ the median efficiency is only $\sim 0.2$, with a large spread.

\begin{figure}[!htp]
\includegraphics[width=\columnwidth]{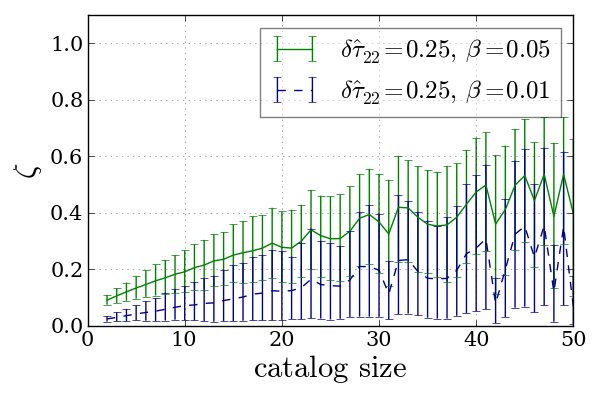}
  \caption{Growth of the efficiency with the number of sources per catalog, this time for a 25\% deviation in $\tau_{22}$, again for maximum tolerable false alarm probabilities of $\beta = 0.05$ and $\beta = 0.01$. As in Fig.~\ref{fig:efficiencies}, medians and 95\% confidence intervals for $\zeta$ are shown, obtained from combining simulated sources into catalogs in many different ways.}
\label{fig:dtau22_25pc}
\end{figure}

One may then wonder how large a deviation in $\tau_{22}$ needs to be before it becomes detectable with good efficiency, still assuming a few tens of sources per catalog. In Fig.~\ref{fig:dtau22_25pc}, we show the evolution of median efficiencies and 95\% confidence intervals for the case where $\delta\hat{\tau}_{22} = 0.25$. Here the efficiencies rise more steeply with the number of detections available, with the median efficiency for $\beta = 0.05$ reaching $\sim 50\%$, albeit still with a considerable spread.

\begin{figure*}[!htp]
  \begin{tabular}{cc}%{@{}ccc@{}}
    \includegraphics[width=.5\textwidth]{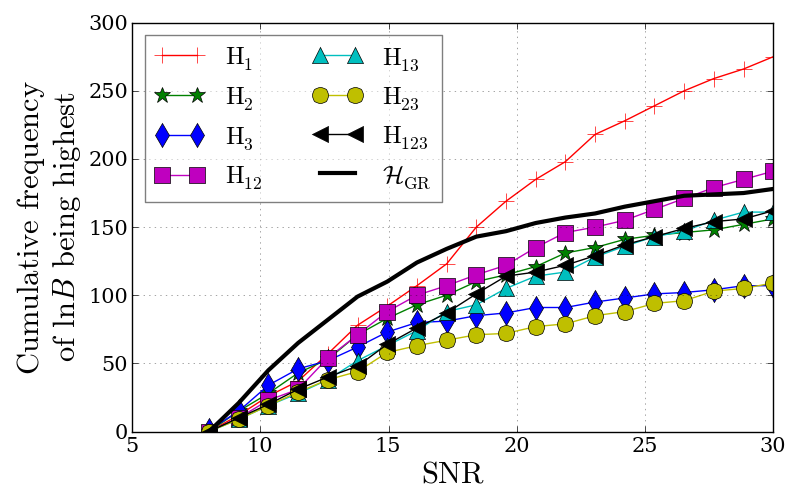} &
    \includegraphics[width=.5\textwidth]{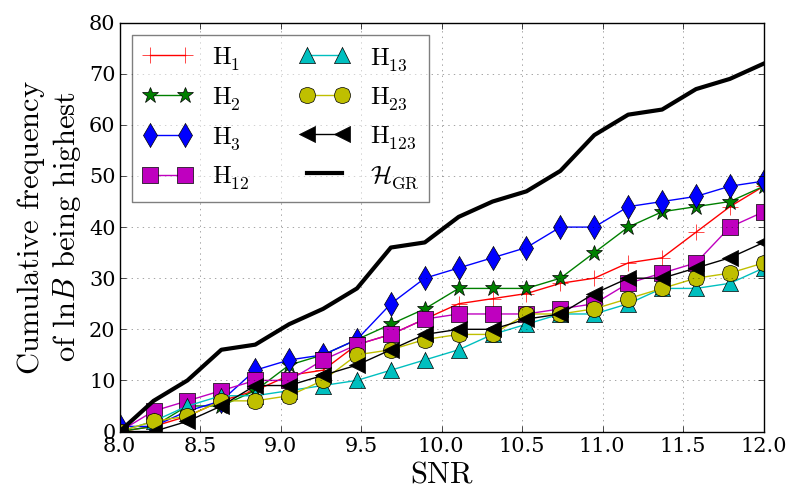}
  \end{tabular}
  \caption{Left: The cumulative number of times that a given hypothesis (any of the $H_{i_1 i_2 \ldots i_k}$, or $\hyp_{\rm GR}$) has the largest Bayes factor against noise ($B^{i_1 i_2 \ldots i_k}_{\rm noise}$, or $B^{\rm GR}_{\rm noise}$), for single sources up to an SNR of 30. Right: The same, but focusing on SNRs up to 12.}
\label{fig:cumulativefrequencies}
\end{figure*}

We see that, by combining information from multiple sources, we greatly improve our ability to use ringdown signals observed by ET in testing the no-hair theorem. However, the advantages of TIGER are not limited to this. The use of multiple sub-hypotheses $H_{i_1 i_2 \ldots i_k}$ also has a significant impact in finding a deviation from GR, as illustrated in Fig.~\ref{fig:cumulativefrequencies}. Here we arrange simulated sources in order of increasing SNR, and we consider the Bayes factors $B^{i_1 i_2 \ldots i_k}_{\rm noise}$ and $B^{\rm GR}_{\rm noise}$ for the hypotheses $H_{i_1 i_2 \ldots i_k}$ and $\hyp_{\rm GR}$ against the noise-only hypothesis, respectively. In particular, what is plotted is the cumulative number of times that the Bayes factor against noise for a particular hypothesis is the largest. We do this for the case where $\delta\hat{\omega}_{22} = 0.1$, so that the correct hypothesis is $H_1$. For SNRs up to $\sim 18$, we see that the GR hypothesis dominates. Going to higher SNRs, the correct hypothesis comes out on top the largest number of times. Even so, \emph{incorrect hypotheses often dominate}. For example, the number of times that the incorrect hypothesis $H_{12}$ has the largest Bayes factor against noise is not significantly lower than the number of times that $H_1$ has the largest Bayes factor. As the right hand panel in the Figure shows, at SNRs between 8 and 12, the hypothesis $H_3$ tends to be the most dominant after $\hyp_{\rm GR}$, yet it does not even involve $\delta\hat{\omega}_{22}$, where the GR violation occurs! Note also that in the latter SNR range, the least parsimonious hypothesis, $H_{123}$, does particularly badly.

\begin{figure}[!htp]
\includegraphics[width=\columnwidth]{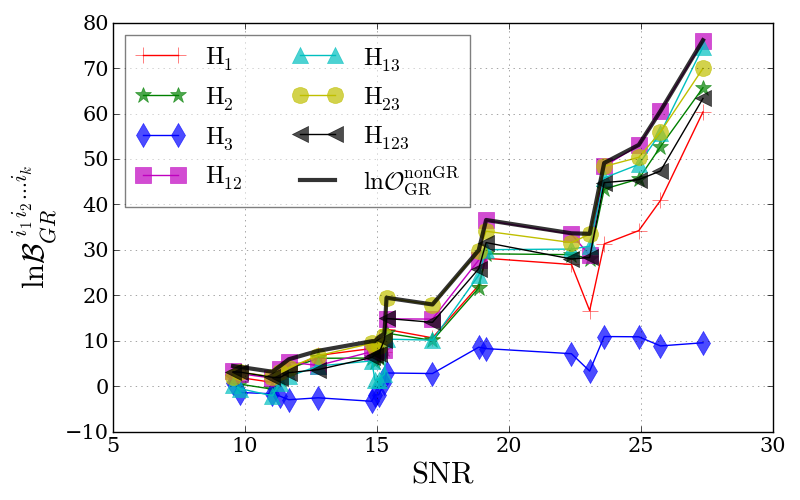}
\caption{The progression of \emph{combined} log Bayes factors within an example \emph{catalog} of 20 sources. Shown are $\ln \mathcal{B}^{i_1 i_2 \ldots i_k}_{\rm GR}$, as well as the log odds ratio $\ln\mathcal{O}^{\rm modGR}_{\rm GR}$, with an increasing number of sources (sorted by SNR), for the case where the injections have $\delta\hat{\omega}_{22} = 0.1$.}
\label{fig:bayesfactors}
\end{figure}

The above pertained to single sources. In Fig.~\ref{fig:bayesfactors} we consider, for an example \emph{catalog} of sources, the evolution of the \emph{combined} Bayes factors
\be
\mathcal{B}^{i_1 i_2 \ldots i_k}_{\rm GR} = \prod_{A = 1}^\mathcal{N} {}^{(A)}B^{i_1 i_2 \ldots i_k}_{\rm GR},
\ee
as well as $\ln\mathcal{O}^{\rm modGR}_{\rm GR}$, as information from more and more detections is added; the sources are arranged in order of increasing SNR. We can make two observations:
\begin{itemize}
\item The hypothesis $H_{123}$ where all the parameters $\{\delta\hat{\omega}_{22}, \delta\hat{\omega}_{33}, \delta\hat{\tau}_{22}, \}$ are left free does not dominate the log odds ratio, and indeed is deprecated compared with some of the other sub-hypotheses. This illustrates how one can suffer significant loss in discriminatory power if the non-GR model is insufficiently parsimonious, \emph{i.e.}~has more free parameters compared with the number of additional parameters that is actually needed. TIGER does not have this problem.
\item The correct hypothesis, in this case $H_1$, is \emph{also} not necessarily the dominant one. Indeed, it can happen that detector noise obscures the true nature of the GR violation so that some other hypothesis (in this example $H_{12}$) ends up on top. However, what is unlikely to happen is that the noise makes a non-GR signal look like a GR one. In a situation where most signals are weak, one should use TIGER with as many testing parameters $\{\delta\xi_1, \delta\xi_2, \ldots, \delta\xi_{N_T}\}$ as is computationally feasible.
\end{itemize}

So far we have considered situations where GR violations are present, and we have studied how well one would be able to find them using TIGER, depending on the size of the violations and the number of detections available. In the next section we consider a scenario where the measured log odds ratio is consistent with GR.

\section{Constraining the free parameters}
\label{sec:PDFs}

\begin{figure*}[!htp]
\centering
  \begin{tabular}{ccc}%{@{}ccc@{}}
    \includegraphics[width=.33\textwidth]{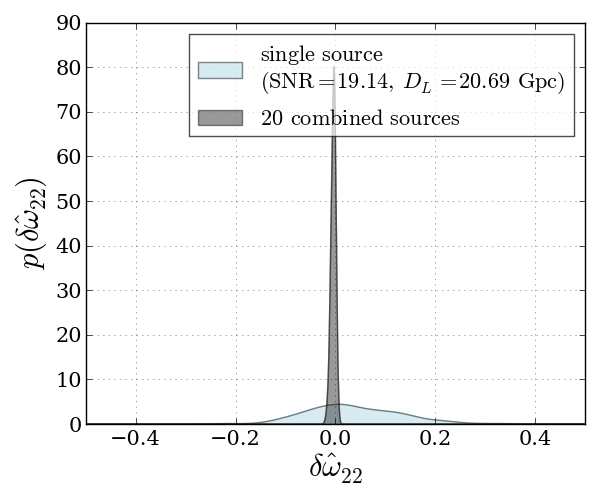} &
    \includegraphics[width=.33\textwidth]{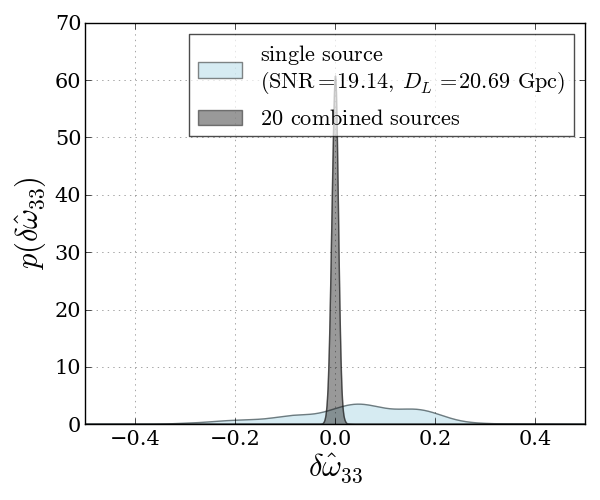} &
    \includegraphics[width=.33\textwidth]{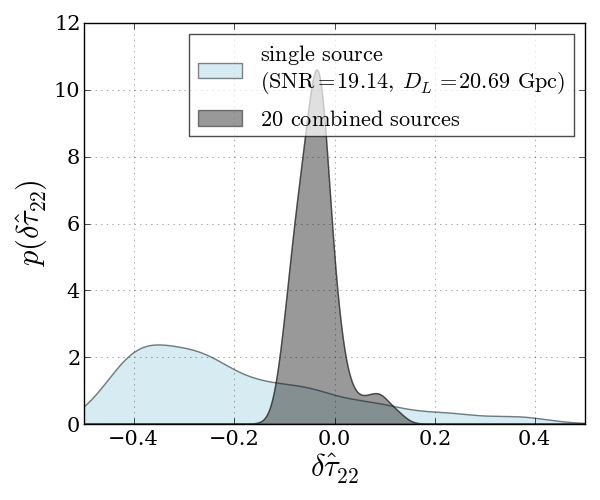} \\
    \includegraphics[width=.33\textwidth]{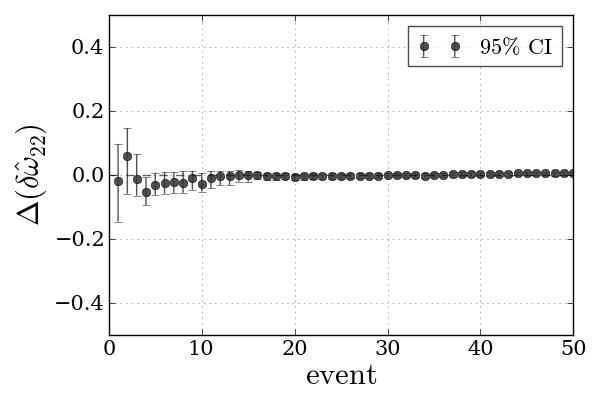} &
    \includegraphics[width=.33\textwidth]{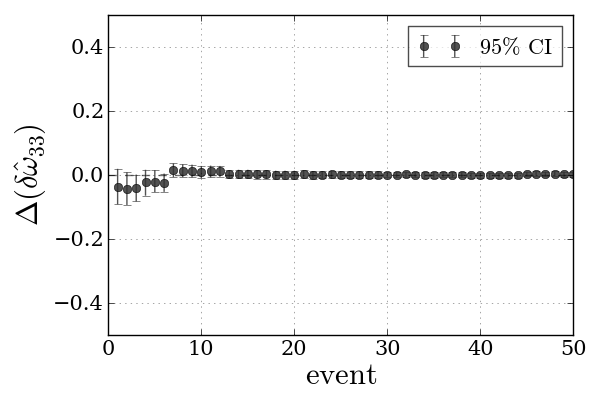} &
    \includegraphics[width=.33\textwidth]{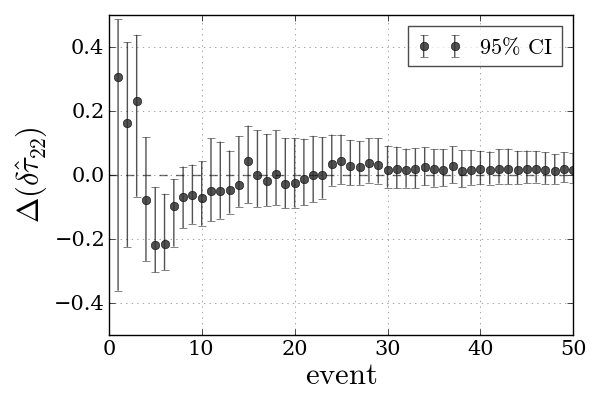}
  \end{tabular}
  \caption{Top panels: Posterior density functions for $\delta\hat{\omega}_{22}$ (left), $\delta\hat{\omega}_{33}$ (middle), and $\delta\hat{\tau}_{22}$ (right), both for a single source at a distance of 20.69 Gpc ($z = 2.47$) with an SNR of 19.14, and for a catalog of 20 sources. Bottom: Evolution of medians and 95\% confidence intervals of PDFs as more and more sources are included.}
\label{fig:PDFs}
\end{figure*}

If, when comparing the measured log odds ratio with some reasonable threshold, there turns out to be no reason to doubt the validity of GR, then one can consider \emph{measuring} the parameters $\delta\hat{\omega}_{22}$, $\delta\hat{\omega}_{33}$, and $\delta\hat{\tau}_{22}$ in order to see what constraints can be put on them. Indeed,  within the Bayesian parameter estimation framework implemented by Veitch and Vecchio \cite{Veitch2008a,Veitch2008b,Veitch2010} that we use here (see also \cite{PEpaper} for comparisons with other Bayesian methods), given a waveform model corresponding to a hypothesis $\hyp$ with parameters $\vec{\lambda}$, the joint \emph{posterior density function} (PDF) is obtained through
\be
p(\vec{\lambda}|\hyp, d, I) = \frac{p(\vec{\lambda}|\hyp, I)\,p(d|\hyp, \vec{\lambda}, I)}{p(d|\hyp, I)},
\ee
where $p(\vec{\lambda}|\hyp, I)$ is the prior distribution of parameters before any measurement has been made, $p(d|\hyp, \vec{\lambda}, I)$ is the likelihood function (\emph{i.e.}~the probability of obtaining the data $d$ given $\hyp$ and parameter values $\vec{\lambda}$), and $p(d|\hyp, I)$ is the prior probability of the data, which can be absorbed in a normalization factor. The probability density function for an individual component $\lambda_1$ of the vector $\vec{\lambda}$ is obtained straightforwardly by marginalizing over all the other parameters:
\be
p(\lambda_1|\hyp, d, I) = \int d\lambda_2 d\lambda_3 \ldots d\lambda_N\,p(\vec{\lambda}|\hyp, d, I).
\ee
If one anticipates $\lambda_1$ to be the same for all detections $d_1, d_2, \ldots, d_\mathcal{N}$ then one can combine PDFs across sources through
\ba
&& p(\lambda_1| \hyp, d_1, d_2, \ldots, d_\mathcal{N}, I) \nn\\
&& = p(\lambda_1 | \hyp, I)^{1 - \mathcal{N}}\,\prod_{A=1}^\mathcal{N} p(\lambda_1 | \hyp, d_A, I),
\label{PDFmultiplesources}
\ea
as was done in a different context in \emph{e.g.}~\cite{DelPozzo2013}.

In the present context one can obtain PDFs for $\delta\hat{\omega}_{22}$, $\delta\hat{\omega}_{33}$, and $\delta\hat{\tau}_{22}$ by \emph{e.g.}~letting $\hyp$ be $H_1$, $H_2$, and $H_3$, respectively. Now, if there is evidence that GR is violated (because of the measured $\ln\mathcal{O}^{\rm modGR}_{\rm GR}$ being above threshold), then there is no \emph{a priori} reason to assume that the $\{\delta\hat{\omega}_{22}, \delta\hat{\omega}_{33}, \delta\hat{\tau}_{22}\}$ will be the same for all sources. Note that although this is the choice we made for injections in the previous section, even if these additional parameters had been non-constant it would not have been a problem to do \emph{model selection} with multiple sources, since each of the Bayes factors ${}^{(A)}B^{i_1 i_2 \ldots i_k}_{\rm GR}$ only gauge whether the hypothesis $H_{i_1 i_2 \ldots i_k}$ is more probable than $\hyp_{\rm GR}$. In doing \emph{parameter estimation} one has to be more careful.

On the other hand, suppose that there is no evidence of GR being incorrect; \emph{i.e.}, the measured $\ln\mathcal{O}^{\rm modGR}_{\rm GR}$ is well below threshold. Then one can expect that $\{\delta\hat{\omega}_{22}, \delta\hat{\omega}_{33}, \delta\hat{\tau}_{22}\}$ are all constant, namely $\delta\hat{\omega}_{22} = \delta\hat{\omega}_{33} = \delta\hat{\tau}_{22} = 0$, and it makes sense to combine PDFs from multiple sources as in Eq.~(\ref{PDFmultiplesources}). In turn, we let $\hyp$ be $H_1$, $H_2$, and $H_3$, and compute marginalized PDFs for $\delta\hat{\omega}_{22}$, $\delta\hat{\omega}_{33}$, and $\delta\hat{\tau}_{22}$, respectively.

Results are shown in Fig.~\ref{fig:PDFs}. In the top panels, we consider PDFs both for an example single source at $D_{\rm L} = 20.69$ Gpc ($z = 2.47$), and for a catalog of 20 sources. For the single source, the spreads of the PDFs are roughly consistent with an extrapolation of the results of Gossan \emph{et al.}~to the given luminosity distance. (For injections and templates with non-spinning progenitors and $D_{\rm L} < 6$ Gpc, we get uncertainties that are in close agreement with theirs.) As expected, the single-source PDFs are quite wide and uninformative, with standard deviations of 0.10, 0.13, and 0.21, respectively. For $\delta\hat{\omega}_{22}$ and $\delta\hat{\omega}_{33}$, with 20 sources the PDFs become strongly peaked (with widths of 0.0051 and 0.0066, respectively), and there is very little bias. 
%Note how for multiple sources the accuracies on $\delta\hat{\omega}_{22}$ and $\delta\hat{\omega}_{33}$ are comparable; with our choice for the range of injected mass ratios ($0.3 <  q < 1$, or $0.18 \lesssim \nu < 0.25$) there will be sources for which $A_{33} > A_{22}$ (see Fig.~1 in \cite{Gossan2012}). 
For $\delta\hat{\tau}_{22}$ the combined PDF is not only wide (with a standard deviation of 0.048), it also shows some bias (although the correct value of zero is within its support). In the bottom panels of the figure, we show the evolution of medians and 95\% confidence intervals for the combined PDF as more and more detections are added. We see that even for $\delta\hat{\tau}_{22}$, the 95\% confidence interval shrinks to $\sim 0.1$ when $\sim 40$ sources are at our disposal. Hence there is a clear advantage in computing PDFs using all available detections.

\section{Conclusions}
\label{sec:Conclusions}

We have revisited the problem of testing the no-hair theorem using ringdown signals that will be seen by Einstein Telescope. In previous work \cite{Gossan2012}, it was shown how deviations of up to 10\% in the ringdown mode frequencies $\omega_{22}$, $\omega_{33}$ and the damping time $\tau_{22}$ could be observed out to distances of $\sim 6$ Gpc, both through parameter estimation and model selection. Here we used the TIGER framework that was originally developed to test general relativity with stellar mass binary inspiral signals in second-generation detectors \cite{Li2012a,Li2012b,VanDenBroeck2013,Agathos2013,Agathos2014}. In this model selection scheme, parameterized deviations are introduced in the waveforms, and multiple auxiliary hypotheses are tested corresponding to all subsets of the extra free parameters. Information from multiple sources can trivially be combined. A log odds ratio $\ln\mathcal{O}^{\rm modGR}_{\rm GR}$ is computed, which compares the probability that one or more of the auxiliary hypotheses are correct with the probability that GR is the right theory. Given the expected distribution of $\ln\mathcal{O}^{\rm modGR}_{\rm GR}$ in the case that GR is correct, violations of GR are searched for by checking whether the measured log odds ratio is above a threshold set by a pre-determined maximum false alarm probability. If this is not the case then there is no reason to doubt GR, and one can calculate bounds on the free parameters, again combining information from all available sources.

Ringdown signals from black holes with masses in the range $500-1000\,M_\odot$ can result from coalescences of intermediate-mass binary black holes, but such events may be rare \cite{Miller2004,Fregeau2006,Amaro-Seoane2006,Amaro-Seoane2010,Gair2011}. On the other hand, they can be seen with ET out to redshifts of $z \gtrsim 5$. We have shown that with $\mathcal{O}(10)$ sources and using the TIGER framework, deviations of the same size as the ones considered in \cite{Gossan2012} can be seen, but for sources at distances up to 50 Gpc.

Our work illustrates how TIGER is not tied to any particular waveform model (nor even any particular type of source). It is well-suited to the regime of low signal-to-noise ratios due to its use of multiple sub-hypotheses, which increases the chance of finding a GR violation. Because of detector noise, the correct hypothesis (or for that matter, the most inclusive hypothesis) may not yield the largest contribution to the log odds ratio, but it is unlikely that noise will make a GR-violating signal look like one that is in accordance with GR. For concreteness we only considered possible deviations in $\{\omega_{22}, \omega_{33}, \tau_{22}\}$ (as was also done in \cite{Gossan2012}), leading to seven auxiliary hypotheses in the TIGER framework, but in reality one should include as many sub-hypotheses as is feasible and modeling allows. 

TIGER offers an effective way of finding very generic violations of GR. As shown in the context of compact binary coalescence, it can uncover deviations that are not included in any of the waveform models associated with the sub-hypotheses $\hyp_{i_1 i_2 \ldots i_k}$ \cite{Li2012a,Li2012b}. We fully expect the same to be true for ringdown; an example could be the appearance of modes with spin weights different from $-2$, as in the case of a black hole in certain $f(R)$ theories that are dynamically equivalent to Einstein-Proca theory \cite{Pani2012,Vitagliano2010,Buchdahl1979}. An explicit demonstration is left for future work. On the other hand, if the signal waveform is not among any of the template models, then fundamental bias can make it difficult to reliably pinpoint the underlying nature of the violation \cite{Yunes2009, DelPozzo2011,Li2012a,Vitale2014,Sampson2014}. Also for ringdown, this remains an open problem.

When the log odds ratio does not indicate a violation of GR, upper limits can be put on deviations in the extra free parameters. A single source at large distance ($> 10$ Gpc) may only give weak bounds and could show considerable bias. On the other hand, with $\mathcal{O}(10)$ sources, deviations are well-constrained even for the parameter $\tau_{22}$, for which no meaningful bounds can be obtained with a single source at SNR $\sim 20$.

In this study we deliberately restricted attention to a black hole mass range for which the preceding inspiral signal can not be seen, but the dominant ringdown mode is visible. In reality one would also expect lighter systems to be seen, for which one would want to utilize information from the inspiral and merger regimes as well. Given appropriate GR waveform models (as are likely to become available on the timescale of ET) it should be possible to put extremely stringent restrictions on GR violations by using the thousands of stellar-mass binary coalescence events that ET will plausibly observe. However, as we have shown, even events where only the ringdown can be accessed will separately allow for interesting tests of the strong-field dynamics of GR.

Finally, as found in \cite{Gossan2012} for the case of \emph{single} systems with $M \sim 10^6 \, M_\odot$, \emph{e}LISA will be able to perform tests of the no-hair theorem at a comparable level of accuracy as ET with $M \sim 10^3\,M_\odot$. Since the detection rate for such sources with \emph{e}LISA may be in the order of tens per year \cite{Amaro-Seoane2012} (\emph{i.e.}~what we assumed for ET in this paper), results from TIGER, including the combining of information from multiple sources, should also be similar. Detailed investigations are left for future work.

\section*{Acknowledgements}

MA, JM, CVDB and JV were supported by the research programme of the Foundation for Fundamental Research on Matter (FOM), which is partially supported by the Netherlands Organisation for Scientific Research (NWO). JV was also supported by STFC grant ST/K005014/1. BSS was supported by STFC grants ST/L000962/1 and ST/L000342/1. It is a pleasure to thank E.~Berti, V.~Cardoso, and I.~Kamaretsos for useful comments and suggestions. 

%
% References
%

\end{document}